\documentclass[prl,twocolumn,letterpaper,showpacs,groupedaddress,superscriptaddress,nofootinbib,floatfix,preprintnumbers,tightenlines]{revtex4}
\usepackage{hyperref,amssymb,amsmath,graphicx,xcolor,cancel}
\usepackage{bm}

\def\si{^1 \hskip -0.03in S _0}
\def\siii{^3 \hskip -0.025in S _1}
\def\diii{^3 \hskip -0.03in D _1}

\begin{document}

\begin{figure}[!t]
\vskip -1.1cm
\leftline{
\includegraphics[width=3.0 cm]{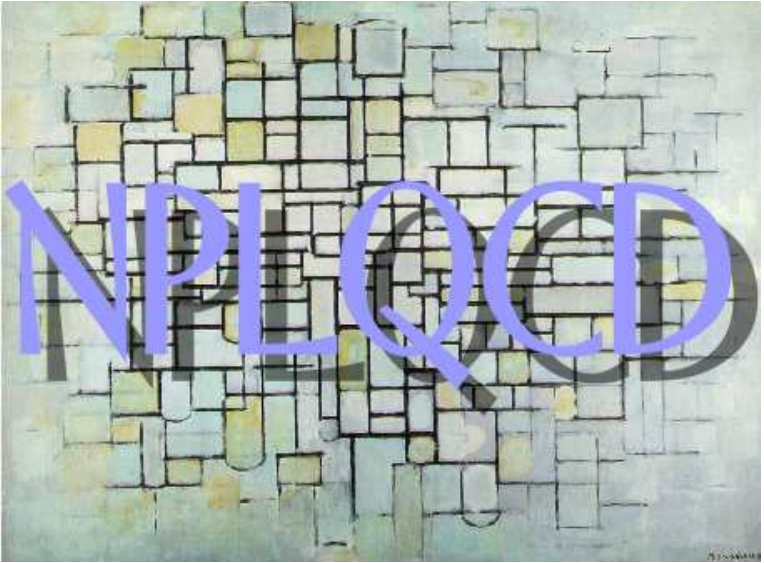}}
\vskip -0.5cm
\end{figure}

\title{Unitary Limit of Two-Nucleon Interactions in Strong Magnetic Fields}

 \author{William Detmold} \affiliation{
Center for Theoretical Physics, 
Massachusetts Institute of Technology, 
Cambridge, MA 02139, USA}
 
\author{Kostas~Orginos}
\affiliation{Department of Physics, College of William and Mary, Williamsburg,
  VA 23187-8795, USA}
\affiliation{Jefferson Laboratory, 12000 Jefferson Avenue, 
Newport News, VA 23606, USA}

\author{Assumpta~Parre\~no}
\affiliation{Dept. d'Estructura i Constituents de la Mat\`eria. 
Institut de Ci\`encies del Cosmos (ICC),
Universitat de Barcelona, Mart\'{\i} Franqu\`es 1, E08028-Spain}

\author{Martin J. Savage}
\affiliation{Institute for Nuclear Theory, University of Washington, Seattle, WA 98195-1550, USA}

 \author{Brian C. Tiburzi} 
\affiliation{ Department of Physics, The City College of New York, New York, NY 10031, USA }
\affiliation{Graduate School and University Center, The City University of New York, New York, NY 10016, USA }
\affiliation{RIKEN BNL Research Center, Brookhaven National Laboratory, Upton, NY 11973, USA }

\author{Silas~R.~Beane} 
\affiliation{Department of Physics,
	University of Washington, Box 351560, Seattle, WA 98195, USA}

\author{Emmanuel~Chang}
\affiliation{Institute for Nuclear Theory, University of Washington, Seattle, WA 98195-1550, USA}

\collaboration{NPLQCD Collaboration}

\date{\today}

\preprint{INT-PUB-15-044}
\preprint{NT@UW-15-10}
\preprint{NSF-KITP-15-119}
\preprint{MIT-CTP-4704}


\begin{abstract}
Two-nucleon systems are shown to exhibit large scattering lengths in strong magnetic fields at unphysical quark masses, and 
the trends toward the physical values indicate that such  features may exist in nature.
Lattice QCD calculations 
of the energies of one and two nucleons systems
are performed 
at pion masses of $m_\pi\sim 450$ and 806 MeV in 
uniform, time-independent magnetic fields of strength 
 $|{\bf B}| \sim 10^{19}$---$10^{20}$ Gauss
to determine the response of these hadronic systems to large magnetic fields. 
Fields of this strength may exist inside magnetars  and in 
peripheral relativistic heavy ion collisions, 
and the unitary behavior at large scattering lengths may have important consequences for these systems.
\end{abstract}

\maketitle


In most physical situations, external electromagnetic (EM) 
fields have only small effects on hadronic and nuclear systems, 
whose structure and dynamics are dominated by the internal strong interactions arising from
Quantum Chromodynamics (QCD) and  internal EM interactions. 
However, there are specific situations involving extremely large EM fields,  created either naturally in astrophysical environments or in particle colliders, for which the effects of external fields are 
important.
In \emph{magnetars}, high magnetic field rotating neutron stars \cite{Duncan:1992hi},  surface magnetic fields are observed up to 
${\cal O}(10^{14})\ {\rm Gauss}$ (for reviews, see e.g.~Ref.~\cite{Harding:2006qn,Camilo:2008nature}), 
and it is conjectured that interior magnetic fields reach up to ${\cal O}(10^{19})\ {\rm Gauss}$ \cite{Broderick:2000pe}. 
In  heavy ion collisions, the  currents produced by relativistic nuclei lead to  large magnetic fields within the projectiles, 
particularly during (ultra-)peripheral collisions~\cite{McLerran:2013hla}. 
It is estimated that fields of ${\cal O}(10^{19})\ {\rm Gauss}$ are  experienced by the nuclei 
during the femtoseconds of the nuclear crossings~\cite{McLerran:2013hla}. 
Neither of these environments are easy to probe in a controlled way, and the detailed
behavior of nuclei in such fields is an open question. 
As a step toward exploring nuclei in these extreme magnetic environments, 
we present the results of calculations of the effects of  uniform, time-independent magnetic fields
on two-nucleon (as well as two-hyperon) systems performed with the underlying quark and gluon degrees of freedom.
We find that such fields can potentially unbind the deuteron and significantly modify the nucleon-nucleon (NN) interactions in the 
 $\si$ channel. At the unphysical quark masses where the calculations 
 are performed, 
the scattering lengths in both the $\siii$--$\diii$ and $\si$ channels  diverge at particular values of the field strength. Near these values, the low energy dynamics of these systems will become unitary. 
The trends seen towards the physical values of the quark masses suggest that this feature may exist in nature in some of these systems.
The prospect of such resonant behavior in nuclear systems is exciting and it will be important to incorporate this effect into models of magnetars and heavy ion collisions in which the relevant field strengths are probed.

Before presenting the results of our  calculations, 
it is interesting to consider phenomenological expectations for the behavior of such systems.\footnote{Significant effort has been devoted to understanding the nature of the QCD vacuum in strong magnetic fields (see Ref.~\cite{Miransky:2015ava} for a review), but effects specific to hadronic systems are not well studied.}
For small, constant magnetic fields, the responses of the two-nucleon systems beyond their charges
are governed by their magnetic moments if the system has spin, 
and otherwise by their magnetic polarizabilities. 
The deuteron has a magnetic moment such that in a magnetic field in the $z$ direction the $j_z=+1$ component is positively shifted in energy with respect 
to the breakup threshold\footnote{
While the deuteron magnetic moment is positive, it is less than the sum of the neutron and proton magnetic moments.
In a potential model the difference is due to the $d$-state admixture into the predominantly $s$-wave deuteron wave function,
while in NN effective field theories (EFTs) this is 
encapsulated in short-distance two-nucleon interactions with the magnetic field.
}
and so an approach toward unbinding in a magnetic field is plausible. However, higher order responses to the magnetic field may be important,
and at intermediate field strengths, $|e {\bf B}|\sim m_\pi^2$, 
significant deviations from linearity should be anticipated.
In the opposite limit of extremely large magnetic fields, where 
$|e {\bf B}|\gg\Lambda_{\rm QCD}^2$, the asymptotic freedom of QCD implies \cite{Cohen:2008bk} that the eigenstates
evolve towards weakly-interacting up and down quarks
in Landau levels. 
Hence, as the magnetic field tends to infinity, the ground states of dilute systems tend to threshold. When the density of the system is also large and comparable to the scale of Landau orbits, more exotic phases may occur (see Ref.~\cite{Ferrer:2012wa} for a review).

In this work, the numerical technique of Lattice QCD (LQCD) is 
applied to study two-nucleon systems in uniform, time-independent  background magnetic fields, following methods used in  previous studies of the magnetic moments \cite{Beane:2014ora}
and polarizabilities \cite{Chang:2015qxa} 
of nucleons and light nuclei up to atomic number $A=4$.
To understand the phenomenological effects of the strong fields in nuclear environments, a first task is to ascertain the
effects on the two-nucleon interactions. Two particle scattering phase shifts can be accessed in LQCD from the volume dependence of 
two-nucleon energies 
(the L\"uscher method~\cite{Luscher:1986pf,Luscher:1990ck}), but here a simpler approach is undertaken in which only the bound states
of the two-nucleon sector are addressed.\footnote{At unphysically large values of the light quark masses, both the deuteron and dineutron are bound, as are various two baryon hypernuclei \cite{Beane:2012vq}.}
The primary goal of these calculations is to investigate how the binding energies of the  
two-nucleon states respond to applied magnetic fields.

 LQCD calculations were performed using two ensembles of 
gauge-field configurations generated with a clover-improved fermion 
action~\cite{Sheikholeslami:1985ij} and the L\"uscher-Weisz gauge 
action~\cite{Luscher:1984xn}. The first ensemble had $N_f=3$ degenerate light-quark flavors with
masses tuned to the physical strange quark mass, producing a 
pion of mass $m_\pi\sim 806~{\rm MeV}$, and used a volume of $L^3\times T=32^3\times48$.
The second ensemble used  $N_f=2+1$ quark flavors with the same 
strange quark mass and degenerate up and down quarks with masses corresponding to a pion mass of 
$m_\pi\sim 450~{\rm MeV}$ and a volume of  $L^3\times T=32^3\times96$. Both 
ensembles had a gauge coupling of $\beta=6.1$, corresponding to a 
lattice spacing of $a \sim 0.11~{\rm fm}$. 
The ensembles consisted of $\sim 1,000$ gauge-field configurations at the SU(3) point and $\sim 650$  
configurations at the lighter pion mass, 
each taken at  intervals of 10 hybrid Monte-Carlo trajectories. 
We have extensively studied these ensembles in previous works, and have found that the finite-volume effects to 
both the single nucleon and two-nucleon bound state energies are small~\cite{Beane:2012vq,Beane:2013br}.

As in Refs.~\cite{Beane:2014ora,Beane:2015yha,Chang:2015qxa}, 
background EM ($U_Q(1)$) gauge fields
were implemented through the gauge-links, 
 \begin{eqnarray}
 U^{(Q)}_\mu(x) & = & 
 e^{ i{6\pi Q_q \tilde n\over L^2} x_1 \delta_{\mu,2}} 
 \times 
 e^{ -i{6\pi Q_q \tilde n\over L} x_2 \delta_{\mu,1} \delta_{x_1,L-1}}
 \,,
 \label{eq:Backfield}
 \end{eqnarray}
that give rise to 
uniform magnetic fields along the $x_3$-direction. These
were multiplied onto each QCD 
gauge field in each ensemble (separately for each quark flavor of charge $Q_q$). 
The combined QCD+EM
gauge fields were used to calculate up-, down-, and strange-quark propagators, which were then contracted to form the requisite 
nuclear correlation functions using the techniques of Ref.~\cite{Detmold:2012eu}. 
To ensure periodicity, $\tilde n\in \mathbb{Z}$, and the values  $\tilde n = 0,1,-2,3,4,-6,12$ were used on the SU(3) symmetric ensemble, while
$\tilde n = 0,1,-2,4$ were used on the $m_\pi\sim 450~{\rm MeV}$ ensemble. 
The corresponding field strengths are quantized as $|e{\bf B}|=6\pi |\tilde n|/(a L)^2$, 
giving a field of ${\cal O}(10^{19})$ Gauss for $\tilde n=1$. 
On each configuration, quark propagators were generated from 48 uniformly 
distributed Gaussian-smeared sources for each magnetic field. For further 
details of the production at the SU(3)-symmetric point, see Refs.~\cite{Beane:2012vq,Beane:2013br,Beane:2014ora} and 
in particular, Ref.~\cite{Chang:2015qxa}. Analogous methods were used for the light mass ensemble.
 
\begin{figure}[!ht]
\hspace*{-0.5cm}\includegraphics[width=1.1\columnwidth]{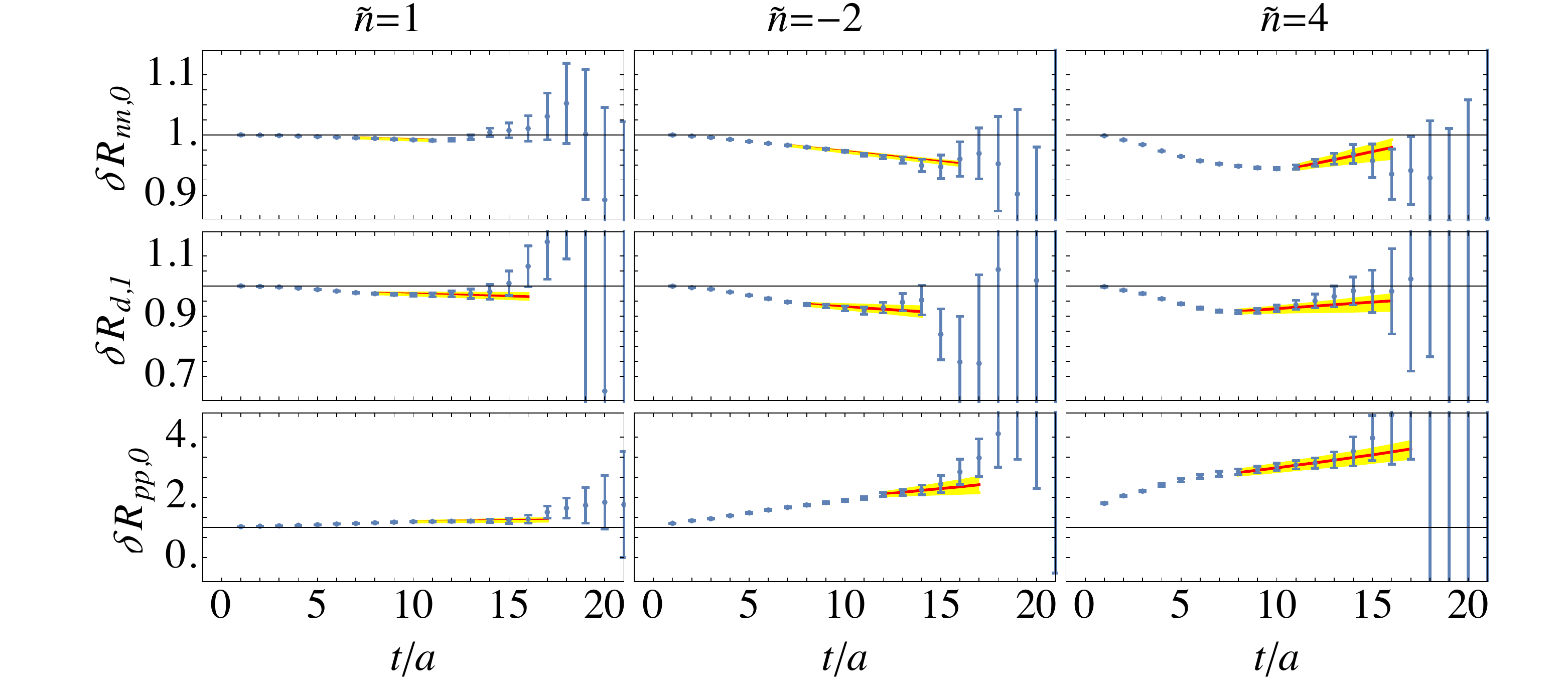}
	\caption{
	Correlator ratios defined in Eq.~(\protect\ref{eq:ratcorrAminus}) for the $nn$, the $j_z=+1$ deuteron and $pp$ systems for field strengths $\tilde n=1,-2,4$,
	for the $m_\pi\sim450$ MeV ensemble.
		The bands correspond to the exponential fit and its statistical uncertainties associated with the shown fit interval. Systematic uncertainties from the choice of fit range are separately assessed.
	\label{fig:corrs}
}
\end{figure}

This work focuses on the dineutron, the diproton, and the maximal  $|j_z|=j=1$  spin state of the deuteron, 
all of which remain isolated, sub-threshold states in the presence of a magnetic field. 
The $I_z=j_z=0$ neutron-proton systems with $(j=1;I=0)$ and $(j=0;I=1)$ mix in a magnetic field and 
have been considered previously in Ref.~\cite{Beane:2015yha}
to determine the cross section for the radiative capture process $np\rightarrow d\gamma$. 
States with the quantum numbers of  $h=n,p,nn,\ pp,\ d_{|j_z|=1}$ are accessed from 
correlation functions $C_h(t; {\bf B})=\langle 0 | \chi_h(t) \overline{\chi}_h(0)| 0 \rangle_{\bf B}$
computed in the presence of the background magnetic field ${\bf B}$  from source and sink interpolating operators with the requisite 
quantum numbers, as discussed in detail in Ref.~\cite{Chang:2015qxa}. 
Representative correlation functions for the heavier mass ensemble can be found in Ref.~\cite{Chang:2015qxa} for
each  hadron/nucleus and background magnetic field. 
Ratios of these correlation functions to those without the magnetic 
field, $R_{h}(t; {\bf B})\equiv C_h(t; {\bf B})/C_h(t; {\bf 0})$,  
are also shown in Ref.~\cite{Chang:2015qxa}, 
and are used to extract the  magnetic moments and polarizabilities 
of the respective systems. 
For the $m_\pi\sim450$ MeV ensemble, the ratios behave in a qualitatively similar manner
and the signals are of comparable quality. 
As the central focus of this study is on the difference between 
the effect of the field on the two-nucleon systems and on the nucleons in isolation,
the further ratios
\begin{eqnarray}
\delta R_{{\cal A}}(t;{\bf B}) & = & 
{ 
R_{{\cal A}}(t; {\bf B}) 
\bigg/
	\prod\limits_{h\in {\cal A}} R_{h}(t; {\bf B}) } 
	\ \ \ ,
\label{eq:ratcorrAminus}
\end{eqnarray}
are of primary importance. In this expression, ${\cal A}$ refers to the composite system and the product is over its constituent nucleon correlator ratios (e.g., for ${\cal A}=d_{j_z=+1}$ the contributions are from $p^\uparrow$ and $n^\uparrow$).
The late time exponential decay of this ratio is dictated  
by the binding energy of the system in the presence of the field~\cite{Chang:2015qxa},
\begin{eqnarray}
\delta R_{{\cal A}}(t;{\bf B})
& \stackrel{t\to\infty}{\longrightarrow} & 
 Z_{{\cal A}}({\bf B})e^{- \left(\delta E_{{\cal A}}({\bf B}) - \sum\limits_{h\in{\cal A}} \delta E_{h}({\bf B}) \right) t }
\,.
\label{eq:ratcorrAminusasymp}
\end{eqnarray}
Fig.~\ref{fig:corrs}, shows these ratios for the $m_\pi\sim450$ MeV ensemble along with the results of single exponential 
fits to time ranges in which the individual correlation functions entering the ratios are consistent with single exponential behavior. 
As discussed in Ref \cite{Chang:2015qxa}, multiple different interpolating operators are investigated for each state in this study and the resulting differences are used to gauge, in part, the systematic uncertainty.
In the figures below, we focus on a particular set of interpolating
operators for clarity but have verified that other choices of interpolators 
provide consistent results.
The analogous results for the heavier mass ensemble are presented in Ref.~\cite{Chang:2015qxa}.
 \begin{figure}[!ht]
 	\includegraphics[width=0.9\columnwidth]{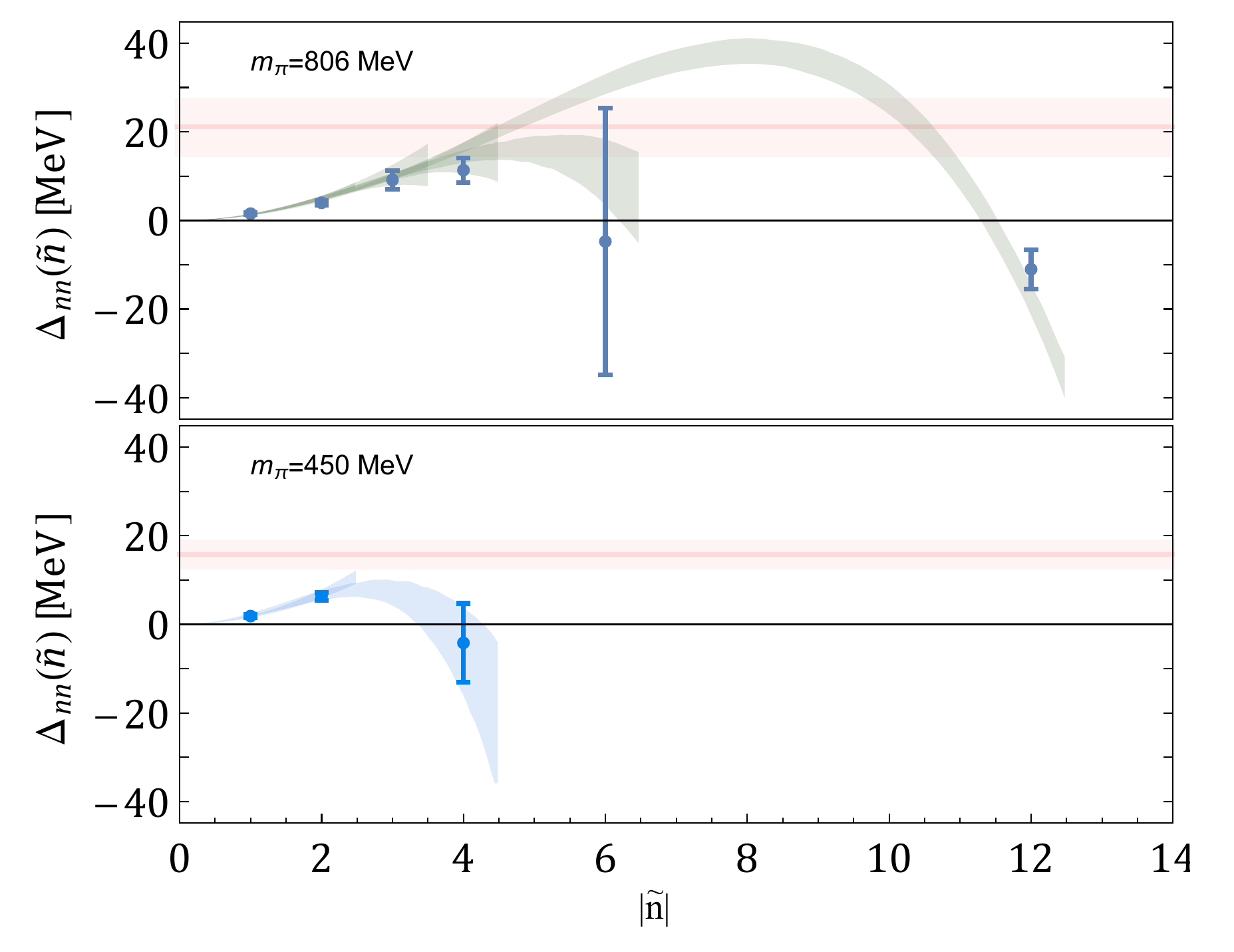}
 	\caption{
	Response of the binding of the dineutron system to applied magnetic fields. The upper panel shows the result at $m_\pi=806$ MeV, while the lower panel is for $m_\pi=450$ MeV.  
	The shaded regions correspond to the 
 		envelopes of  successful fits to the energy shifts using linear and quadratic polynomials in $\tilde n^2$ to data points in the corresponding range indicated by the shaded region. 
		The horizontal bands indicate the binding threshold.
 		\label{fig:nn}
 		}
 \end{figure} 
\begin{figure}[!ht]
	\includegraphics[width=\columnwidth]{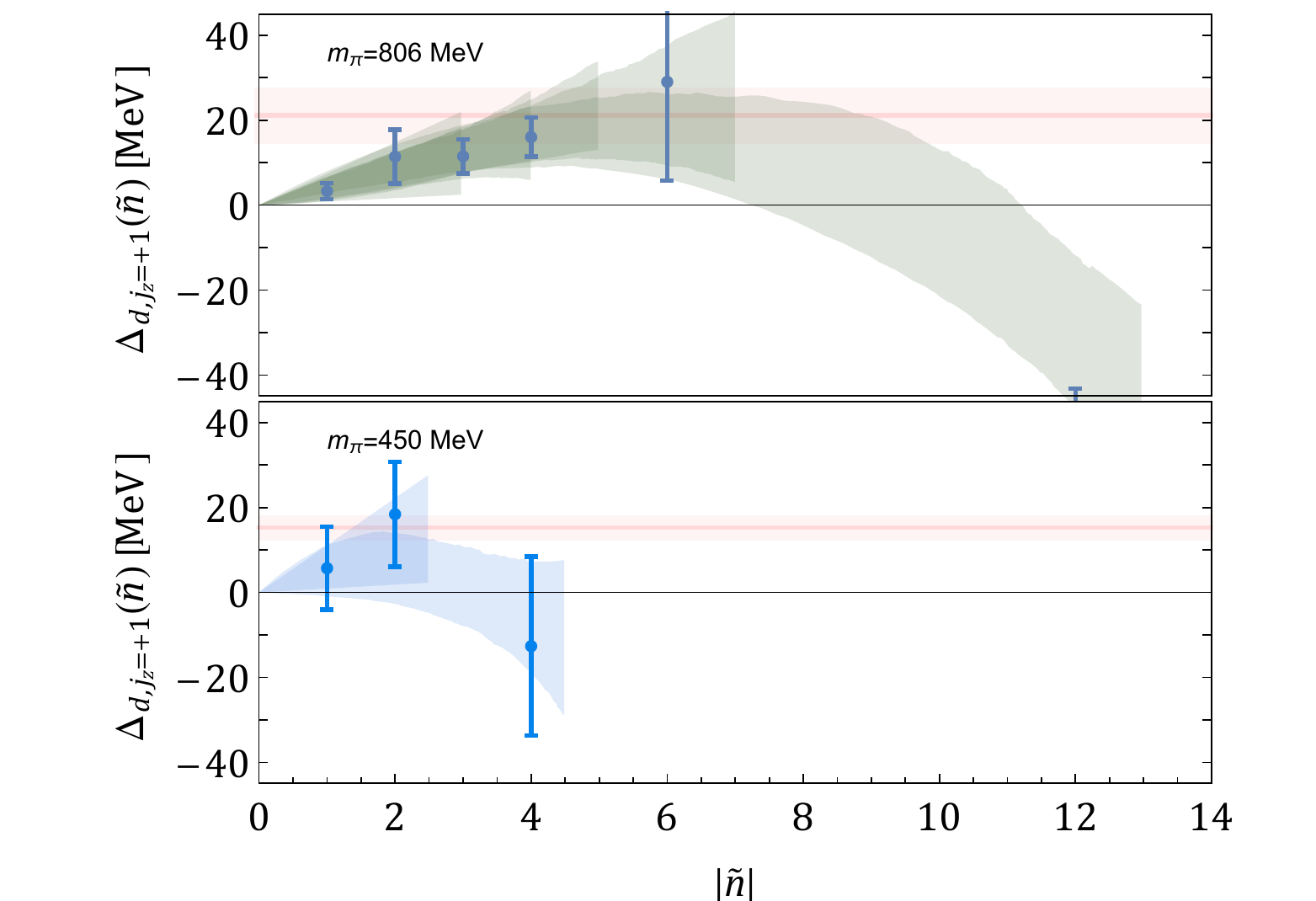}
	\caption{
	Response of the binding of the $j_z=+1$ state of the deuteron to applied magnetic fields.  The shaded regions correspond to the 
		envelopes of successful fits to the energy shifts using polynomials in $\tilde n$ of up to $4^{th}$ ($2^{nd}$) order
		for the $m_\pi=806$ (450) MeV ensemble. The horizontal bands indicate the binding threshold.
		\label{fig:deut}
		}
\end{figure} 

The energy shifts
\begin{equation}
\Delta_{\cal A}(\widetilde n) \equiv \delta E_{{\cal A}}({\bf B}) - \sum\limits_{h\in{\cal A}} \delta E_{h}({\bf B})
\end{equation}
  in the dineutron and deuteron ($j_z=+1$) channels are shown in Figs.~\ref{fig:nn} and \ref{fig:deut}, respectively.
As the strength of the applied magnetic field is increased, the ground state energies of the systems are shifted closer to threshold, and at a given field strength it appears that the states unbind. 
For the deuteron, this behavior is not clearly
resolved  at the lighter mass because of the uncertainties. 
The approach to threshold and subsequent 
turnover is seen at both quark masses 
in the dineutron system,
and the point of minimum binding  decreases as the 
quark mass is lowered, $\tilde n_{nn}^{(\rm max)}\sim 6$ at $m_\pi\sim 806$ MeV and 
$\tilde n_{nn}^{(\rm max)}\sim 3$ at $m_\pi\sim 450$ MeV. 
The dineutron is unbound in nature
and the 
present results suggest that 
magnetic effects would push the system further into the continuum. 
On the other hand, it is possible that the deuteron 
could be unbound by the presence of magnetic fields of strength comparable to those expected in magnetars 
and heavy ion collisions, potentially modifying the dynamics of those systems. 
A particularly interesting aspect of  the behavior in both of these channels is the approach to the unitary regime 
in which the binding energies decrease to zero and consequently the scattering lengths diverge.\footnote{
       It is expected that the range of the interaction (set by hadronic scales) is only weakly affected 
	by the magnetic field, so the volume effects in the two-nucleon systems are not expected to be unmanageable 
	even as the scattering length diverges.
	}
 In atomic physics, such behavior is routinely used to investigate the universal physics that emerges in systems interacting near 
 unitarity~\cite{RevModPhys.82.1225}, but they have not been observed in nuclear
 physics.

\begin{figure}[!t]
	\includegraphics[width=\columnwidth]{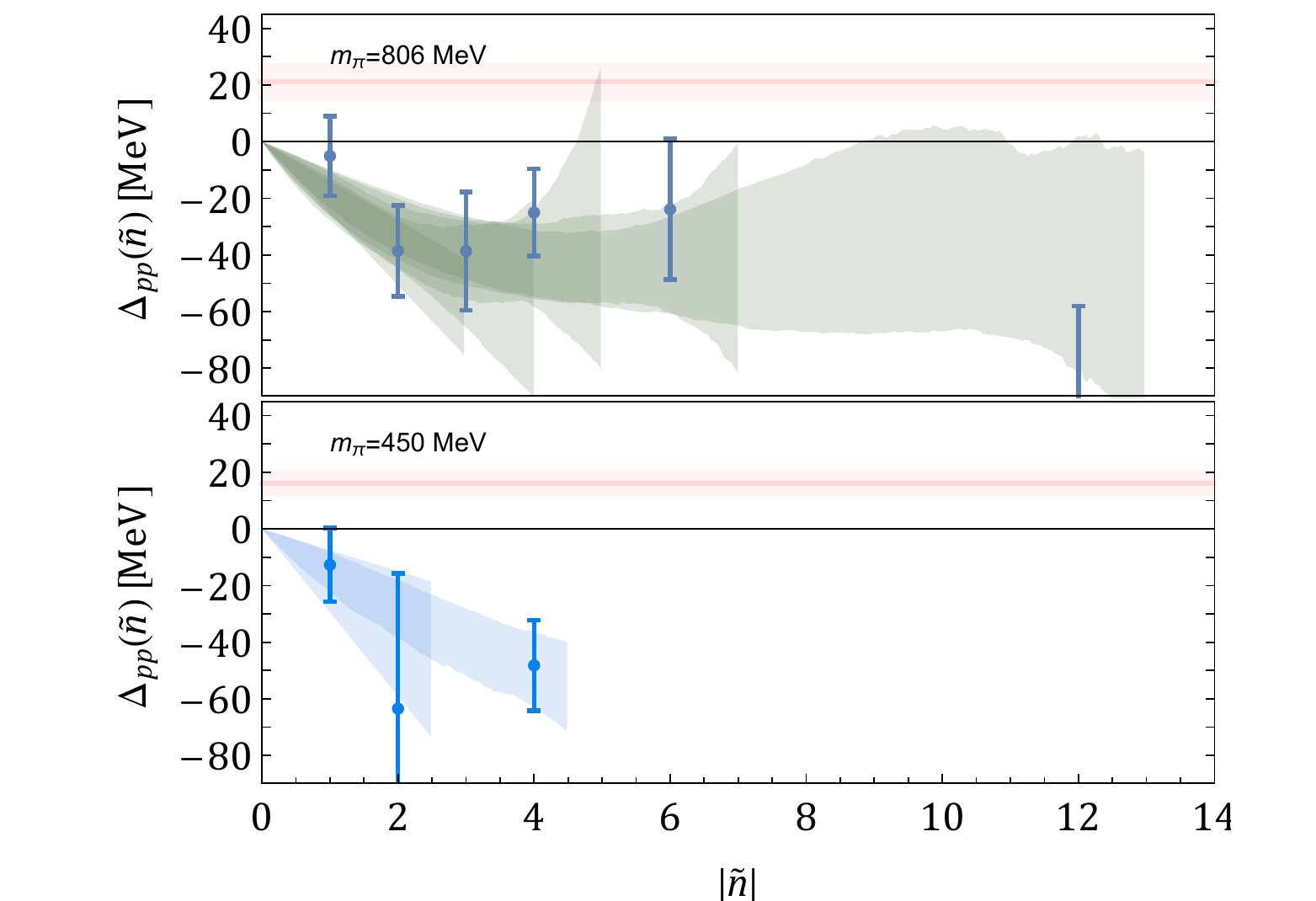}
	\caption{
		Response of the binding of the diproton to applied magnetic fields. The shaded regions correspond to the 
		envelopes of  successful fits to the energy shifts using polynomials in $\tilde n$ of up to $4^{th}$ ($2^{nd}$) order
		for the $m_\pi=806$ (450) MeV ensemble. The horizontal bands indicate the binding threshold.
		\label{fig:pp}
	}
\end{figure} 
\begin{figure}[t]
	\includegraphics[width=0.9\columnwidth]{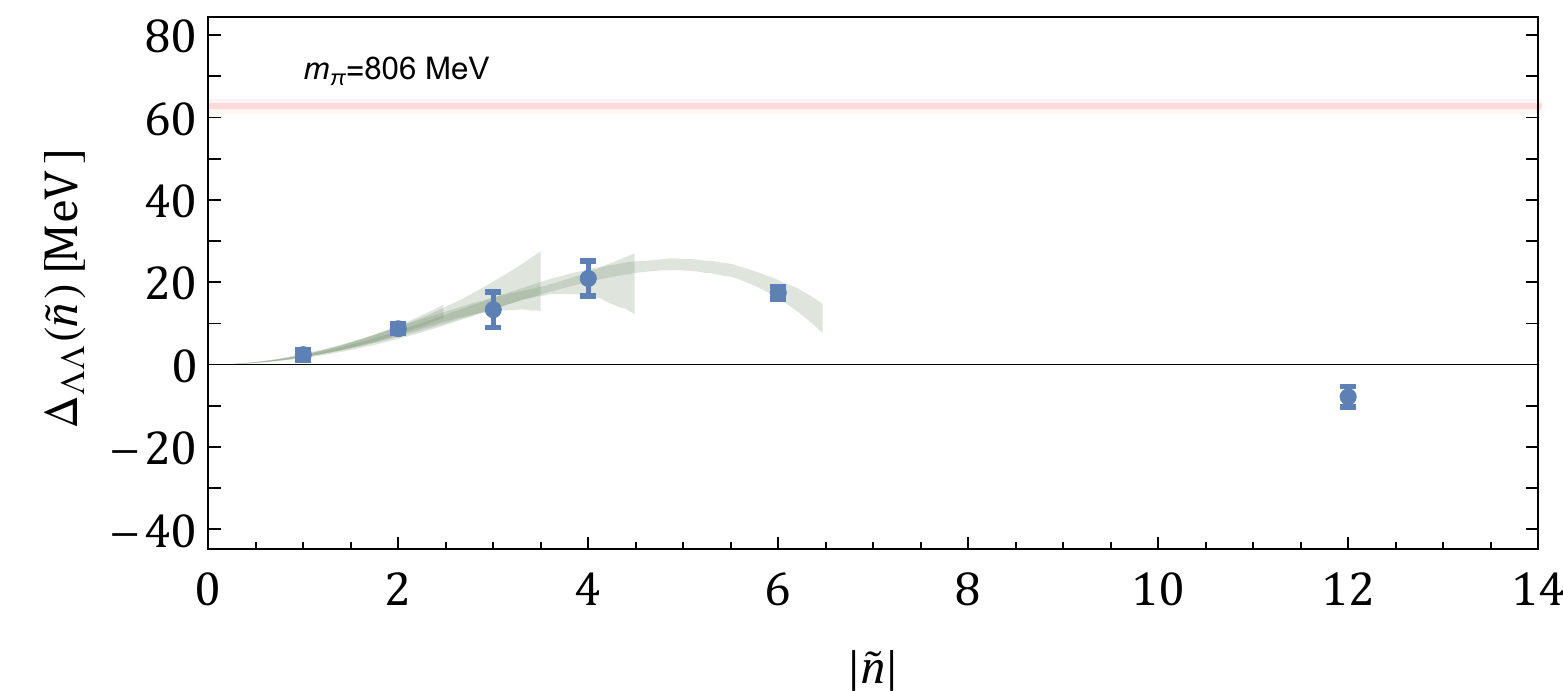}
	\caption{
		The energy splitting in the $H$ dibaryon channel at  $m_\pi=806$ MeV.  The horizontal band indicates the binding threshold.
		\label{fig:lamlam}}
\end{figure}
The energy shifts of the diproton are shown in Fig.~\ref{fig:pp}. 
For this system, the extracted energies are not as cleanly 
determined as for the dineutron, but a trend toward strengthening attraction is 
seen at both quark masses as the field strength increases. 
This is interesting in light of a recent suggestion~\cite{Allor:2006zy} that  the diproton can overcome the Coulomb
repulsion and form a bound state in
sufficiently large  magnetic fields. 
A naive extrapolation of the slope of the shift linearly in $m_\pi^2$ indicates that for a field of $|e {\bf B}|\sim 10^{17}$ Gauss, corresponding to $\tilde n\sim0.01$, the additional attraction
is enough to  bind the 
diproton system. 
While such a result would be interesting, further calculations at lighter quark masses are necessary 
to refine the extrapolation. 

Two baryon systems containing strange quarks have also been investigated. 
Figure \ref*{fig:lamlam} shows the energy splittings of the ground state
in the channel with the quantum numbers of two $\Lambda$-baryons, which contains a deeply bound 
$H$-dibaryon at heavier quark masses~\cite{Beane:2010hg,Inoue:2010es}.
This channel exhibits a slight reduction of the binding energy for intermediate field strengths,
comparable in size to that of the dineutron system,
but  does not exhibit resonant behavior in the range of field strengths that are probed as the binding energy is significantly larger.

\emph{Discussion:}
Having found significant changes in the binding of two-nucleon systems
immersed in strong magnetic fields
at two values of unphysical quark masses, 
it is conceivable that similar modifications occur in nature. 
To solidify this discussion, calculations would need to be performed at or near the physical quark masses and the continuum and infinite volume limits would require careful investigation.\footnote{Based on studies of binding energies on these and  other related ensembles, we are confident that the current calculations do not suffer 
	from large volume or scaling artifacts.}
While the  
responses of these  systems 
can as yet only be estimated at the physical quark masses, the calculated trends provide an interesting  starting point to consider possible consequences.
To this end, it is conjectured that two-nucleon systems will exhibit  unitary behavior, with the deuteron
unbinding in a large magnetic field
and the diproton system becoming bound. On the other hand,
the dineutron will be pushed further into the continuum as the field strength increases. Interestingly, it may be possible to find values of the 
field strength and quark masses where all NN states are at threshold simultaneously, realizing the low energy conformal 
symmetry postulated by Braaten and Hammer \cite{Braaten:2003eu}.
Given the observed behavior of bound states, it is natural to expect that the NN scattering phases shifts and mixing angles will also be modified at a similar level in such fields.
These modifications would be interesting to probe in future LQCD calculations utilizing the L\"uscher method \cite{Luscher:1986pf,Luscher:1990ck} to analyze the spectra of NN systems.

In ultra-peripheral heavy ion collisions, one can speculate that the reduced binding between pairs of nucleons,
along with the reduction in the nucleon mass, will increase the size of each nucleus as they interact with 
the field of the other nucleus. 
Ignoring other potential effects, purely geometrical considerations will result in larger than expected interaction cross-sections that will increase with the collision energy for a given impact parameter and potentially larger fluctuations in collision cross sections. 
However, considering the transient nature of such a collision, and the difference between the internal time-scales 
associated with a rearrangement of the nucleons comprising each nucleus and that of the collision,
more detailed analyses must be performed before even a qualitative 
understanding can be established.
The effects of large magnetic fields in magnetars through the magnetic moments of nucleons and electrons have been considered through a number of model approaches \cite{Broderick:2000pe,Mao:2001cv,Cardall:2000bs,Son:2007ny,PenaArteaga:2011wm,Ferrer:2015wca}, and in some cases lead to 
significant modifications. The more complicated effects from magnetic shifts in binding and hadronic interactions likely also induce significant modifications that deserve further investigation.

\begin{acknowledgments}
	We would like to thank Zohreh Davoudi, Daekyoung Kang and Krishna Rajagopal for several interesting discussions. 
	This research was supported in part by the National Science Foundation under Grant No. NSF PHY11-25915 and
	WD and MJS acknowledge the Kavli Institute for Theoretical Physics for hospitality during completion of this work.
	Calculations were performed using computational resources provided
	by the Extreme Science and Engineering Discovery Environment
	(XSEDE), which is supported by National Science Foundation grant
	number OCI-1053575, NERSC (supported by U.S. Department of
	Energy Grant Number DE-AC02-05CH11231),
	and by the USQCD
	collaboration.  This research used resources of the Oak Ridge Leadership 
	Computing Facility at the Oak Ridge National Laboratory, which is supported 
	by the Office of Science of the U.S. Department of Energy under Contract 
	No. DE-AC05-00OR22725. 
	The PRACE Research Infrastructure resources Curie based in France at the Tr\`es Grand Centre de Calcul and MareNostrum-III based in Spain at the Barcelona Supercomputing Center were also used.
	Parts of the calculations used the Chroma software
	suite~\cite{Edwards:2004sx}.  
	SRB was partially supported by NSF continuing grant PHY1206498 and by U.S. Department of Energy through Grant
	Number DE-SC001347.  
	WD was partially supported by the U.S. Department of Energy Early Career Research Award DE-SC0010495.
	KO was partially supported by the U.S. Department of Energy through Grant
	Number DE- FG02-04ER41302 and through contract number DE-AC05-06OR23177
	under which JSA operates the Thomas Jefferson National Accelerator Facility.  
	The work of AP was supported by the contract
	FIS2011-24154 from MEC (Spain) and FEDER. 
	MJS was supported  by DOE grant No.~DE-FG02-00ER41132.  
	BCT was supported in part by a joint City College of New York-RIKEN/Brookhaven Research Center
	fellowship, a grant from the Professional Staff Congress of the CUNY, and by the U.S. National Science Foundation, under Grant
	No. PHY15-15738.
\end{acknowledgments}
%

\bibliography{bib_feshbach.bib}
\end{document}